\documentclass{aa}

\newcommand{\lta}{\stackrel{<}{\sim}}
\newcommand{\bec}[1]{\mbox{\boldmath $ #1$}}
\begin{document}
\title{Helicity balance and steady-state strength of the dynamo
generated galactic magnetic field}
\author{N.\, Kleeorin \inst{1} \and D.\, Moss \inst{2} \and I.\,
Rogachevskii\inst{1}  \and D.\, Sokoloff\inst{3}} \offprints{D.
Moss} \institute{Department of Mechanical Engineering, Ben-Gurion
University of Negev, POB 653,  84105 Beer-Sheva, Israel\\
\email{nat@menix.bgu.ac.il; gary@menix.bgu.ac.il} \and Department
of Mathematics, University of Manchester, Manchester M13
9PL, UK \\
\email{moss@maths.man.ac.uk} \and Department of Physics, Moscow
State University, Moscow
119899, Russia \\
\email{sokoloff@dds.srcc.msu.su}}
\date{Received 5 June 2000; accepted 4 August 2000}
\abstract{We demonstrate that the inclusion of the helicity flux
in the magnetic helicity balance in the nonlinear stage of
galactic dynamo action results in a radical change in the magnetic
field dynamics. The equilibrium value of the large-scale magnetic
field is then approximately the equipartition level. This is in
contrast to the situation without the flux of helicity, when the
magnetic helicity is conserved locally, which leads to
substantially subequipartition values for the equilibrium
large-scale magnetic field.

}

\maketitle

{\bf ASTRON. ASTROPHYS. 361, L5-L8 (2000)}

\section{Introduction}

The large-scale magnetic fields of galaxies are thought to be
generated by a galactic dynamo due to the simultaneous action of the
helicity of interstellar turbulence and differential rotation (see,
e.g., Ruzmaikin et al. 1988).
The kinematic stage of the galactic dynamo, i.e. the evolution of a weak
magnetic field with negligible influence on the turbulent flows,
seems to be clear, while the nonlinear stage of dynamo evolution is a
topic of intensive discussions (for reviews, see Beck et al. 1996,
Kulsrud 1999). The most contentious issue is  the question of the
equilibrium magnetic field strength at which dynamo action saturates.

A naive viewpoint is that the saturation level for the {\it
large-scale} magnetic field is given by the equipartition between
kinetic energy and the energy of the large-scale magnetic field ${\bf
B}$ (see, e.g., Zeldovich et al.
1983). The motivation is that the equations describing large-scale
dynamo action contain the  mean, but not the total, magnetic field.
This naive outlook leads to models of dynamo generated magnetic fields
which are in basic agreement with the available observational
information.

Vainshtein and Cattaneo (1992) formulated a more sophisticated
argument, suggesting that the equilibrium magnetic field should be
determined by a balance between the kinetic energy and the energy of
the {\it total} magnetic field. The simplest models of dynamo generation
then result in the estimate $b/B \sim Rm^{1/2}$, where $b$ is the
small-scale magnetic field, and the magnetic Reynolds number
$Rm \approx 10^8$  for the interstellar turbulence (or even much larger if
a microscopic diffusivity instead of ambipolar diffusion is used; cf.
Brandenburg \& Zweibel, 1995).
Thus the ideas of Vainshtein and Cattaneo lead to the
conclusion that a dynamo generated large-scale galactic magnetic field
must be negligible in comparison with that observed, and so the generation
of the observed field must be connected with another mechanism. However,
no other general and realistic mechanism for galactic magnetic field
generation is currently available.

The arguments of Vainshtein and Cattaneo do not seem inevitable. For
example, a dynamo generated magnetic field can itself produce helicity,
so the nonlinear effects can even amplify rather than suppress field generation
at the initial stages of nonlinear evolution (Parker 1992, Moss et al.
1999); other suggestions are discussed by, e.g., Beck et al. (1996),
Kulsrud (1999), Field et al. (1999) and Blackman \& Field (1999).
In particular, Blackman \& Field (2000) argue that the $Rm$-dependent
quenching seen in the simulations of Cattaneo \& Hughes (1996) is a
consequence of helicity conservation when using closed or periodic
boundaries, while simulations with open boundaries by Brandenburg \&
Donner (1997) (see also Brandenburg 2000) do not show this effect.

The aim of this letter is to demonstrate that with open boundaries
the scenario of
Vainshtein and Cattaneo
results in basically the same estimate for the equilibrium magnetic
field strength as is given by the naive viewpoint.

The essence of our arguments can be presented as follows. According
to Vainshtein and Cattaneo, the suppression of dynamo action by the
small-scale magnetic field that is generated together with the
large-scale is connected with the magnetic helicity of the
small-scale magnetic field.
Because the total magnetic helicity is an inviscid invariant of
motion, the magnetic helicity of the small-scale magnetic field can
be connected with the magnetic helicity of the large-scale magnetic field.
The governing equation for magnetic helicity has been proposed by
Kleeorin and Ruzmaikin (1982; see the discussion by Zeldovich et al.,
1983), investigated by Kleeorin et al. (1995) for stellar dynamos,
and self-consistently derived by Kleeorin and Rogachevskii (1999).
During nonlinear stages of the dynamo, the $\alpha$-effect is thought
to be determined by the hydrodynamic and magnetic helicities, so a
closed system of equations can be obtained for the evolution of the
magnetic field and the $\alpha$-coefficient (see below, Sect.~2).
This governing system (with helicity locally conserved) leads to magnetic field
behaviour which is consistent with the prediction of Vainshtein and
Cattaneo (we are grateful to M. Reshetnyak, who provided us with the
relevant numerical results, which will be published elsewhere).

We stress that Eq.~(\ref{prime}) takes into account the local
helicity balance at a given point inside the galactic disc $|z|<h$,
$r<R$, where $r, \varphi, z$ are cylindrical coordinates.  However,
the kinematic galactic dynamo is impossible without a turbulent flux
of magnetic field through the surface $|z| = h$ (see, e.g., Zeldovich
et al. 1983, Ch. 11).
It is more than natural to believe that this flux can transport
magnetic helicity to the outside of the disc.
The methods of Kleeorin and Rogachevskii (1999) allow
us to introduce the corresponding term
into the governing equations for the galactic dynamo. We demonstrate
by numerical simulations, and to some extent analytically, that this
term leads to a drastic change  in the magnetic field evolution. Now
the steady-state large-scale magnetic field strength is approximately
in equipartition with the kinetic energy of the interstellar
turbulence.

\section{Equations for magnetic helicity}

Following Kleeorin and Ruzmaikin (1982), we parameterize the
back-reaction of dynamo generated magnetic field in terms of a differential
equation for the $\alpha$-coefficient, using arguments from the magnetic
helicity conservation law.
It is necessary to introduce the large-scale
vector potential ${\bf A}$, small-scale vector potential ${\bf a}$, and the
corresponding representations for the magnetic fields, ${\bf B}$ and
${\bf b}$. We then write the total magnetic field as ${\bf H} = {\bf B}
+ {\bf b}$, and the total vector potential as $\bec{\cal A}={\bf A}+{\bf
a}$,
thus decomposing the fields into mean and fluctuating parts.
The equation for the vector potential $\bec{\cal A}$ follows
from the induction equation for the total magnetic field ${\bf H}$
\begin{eqnarray}
\partial \bec{\cal A} / \partial t = {\bf v \times H}  -
\eta \, {\bf curl \, H} +  \bec{\nabla} \varphi \;,
\label{A2}
\end{eqnarray}
where $ {\bf v} = {\bf V} + {\bf u} ,$ and $ {\bf V} = \langle
{\bf v} \rangle $ is the mean fluid velocity field, $ \eta $ is the
magnetic diffusion due to the electrical conductivity of the fluid,
$ \varphi $ is an arbitrary scalar function. Now we multiply
the induction equation for the total magnetic field ${\bf H}$ by
$ {\bf a} $ and Eq. (\ref{A2}) by $ {\bf b} $, add them and
average over the ensemble of turbulent fields. This yields an equation
for the magnetic helicity $ \chi^{h} = \langle {\bf a} \cdot {\bf b}
\rangle $ in the form
\begin{eqnarray}
\partial \chi^{h} / \partial t + \bec{\nabla} \cdot {\bf F} =
- 2 \langle {\bf u \times b} \rangle \cdot {\bf B} - 2 \eta
\langle {\bf b} \cdot {\bf curl \, b} \rangle \;,
\label{A3}
\end{eqnarray}
where $ {\bf F} = (2/3) {\bf V} \chi^{h} + \langle {\bf a \times}
({\bf u \times B}) \rangle - \eta \langle {\bf a \times} {\bf curl \, b}
\rangle + \langle {\bf a \times} ({\bf u \times b}) \rangle -
\langle {\bf b} \varphi \rangle $ is the flux of magnetic
helicity. The electromotive force for isotropic and homogeneous turbulence
is
\begin{eqnarray}
\langle {\bf u \times b} \rangle = \alpha {\bf B} - \eta_{T}
{\bf curl \, B} \;,
\label{A4}
\end{eqnarray}
where $ \eta_{T} $ is the turbulent magnetic diffusivity, and
it is assumed that $ \alpha $ is the total alpha-effect which
at the nonlinear stage includes both the original hydrodynamical, and the
magnetic, contributions. Note that the magnetic part of the
$\alpha$ effect is proportional to the magnetic helicity, i.e.
$\alpha^h = \chi^{h} / (18 \pi \eta_{T} \rho) $ (see, e.g., Kleeorin
and Rogachevskii, 1999), where $ \rho $ is the density.
The simplest form of the magnetic helicity flux for an isotropic
turbulence is given by $ {\bf F} = {\bf V} \chi^h ,$ where $ {\bf V} $
is the mean fluid velocity, e.g. that of the differential rotation (see
Kleeorin and Ruzmaikin, 1982; Kleeorin and Rogachevskii, 1999).
Thus, the equation for the magnetic part of the $\alpha$ effect
in dimensionless form is given by
\begin{equation}
{{\partial \alpha^h} \over {\partial t}} + {\alpha^h \over T} +
\bec{\nabla} \cdot ({\bf V} \chi^h) = 4 (h/l)^2(R_\alpha^{-1}
{\bf B \cdot curl \, B} - \alpha B^2) \;,
\label{prime}
\end{equation}
(see Kleeorin and Ruzmaikin, 1982), where $l \approx 100 \, pc$ is the
scale of turbulent motions. We adopt here the standard dimensionless
form of the galactic dynamo equation from Ruzmaikin et al. 1988; in
particular, the length is measured in units of the disc thickness $h$,
the time is measured in units of $ h^{2} / \eta_{T} $ and $B$ is
measured in units of the equipartition energy $B_{\rm eq} =
\sqrt{4 \pi \rho} \, u $.  Here $u$ is the characteristic turbulent
velocity in the scale $l$, $\quad \eta_{T} = l u / 3,$ $ \quad T = (1/3)
(l/h)^{2} Rm$ and $R_\alpha = l \alpha_\ast / \eta_{T}$, where $ \alpha^h
$ and $ \alpha $ are measured
in units of $ \alpha_\ast $ (the maximum value of the hydrodynamic
part of the $ \alpha $ effect).
For an axisymmetric dynamo $ \bec{\nabla} \cdot ({\bf V} \chi^h) = 0 .$

When $ \partial \alpha^h / \partial t = 0 $
and $ R_\alpha^{-1} {\bf B \cdot curl \, B} \ll \alpha B^2 $,
Eq.~(\ref{prime}) yields $ \alpha = \alpha^{v} / [1 + (4/3) Rm
B^{2}] $ (see, e.g., Vainshtein and Cattaneo , 1992).
However, the latter equation is not valid for galaxies
because $ \partial \alpha^h / \partial t \gg \alpha^h / T$.
In addition, the condition $ R_\alpha^{-1} {\bf B \cdot curl \, B} \ll
\alpha B^2 $ seems not to be valid for galaxies.

Equation ~(\ref{prime}) has been later reproduced, e.g. by Gruzinov
and Diamond (1995). However, although this equation has never been
included into detailed galactic dynamo calculations, nevertheless its
qualitative properties are more or less clear. Provided that
dissipative losses are taken into account, Eq.~(\ref{prime}) leads to
the same type of behaviour as that obtained
by the {\it ad hoc} prescription of the result of Vainshtein and
Cattaneo (1992), i.e.
the steady state strength of magnetic field is about $B_{\rm eq}
Rm^{-1/2}$ (see, e.g. Field et al., 1999). The real advantage of
Eq.~(\ref{prime}) is the fact that it is derived from first
principles rather  than prescribed {\it ad hoc}.
If the dissipative losses in Eq.~(\ref{prime}) are neglected, the
magnetic field decays for $t\to \infty$.
We stress that Eq.~(\ref{prime}) contains a large factor $4 (h/l)^2
\sim 100$ typically.

Kleeorin and Rogachevskii (1999)
extended the calculations to include a flux of magnetic helicity.
Based on Eq.~(13) of that paper, the approximate relation

\begin{eqnarray}
{{\partial \alpha^h} \over {\partial t}} &=& 4 \biggl({h \over l}
\biggr)^2 [({\bf B \cdot curl \, B} R_\alpha^{-1} - \alpha (B) B^2)
\nonumber \\ &+& {{\partial} \over {\partial z}} (\alpha ^v(z) \phi
(B) B^2 h f_1(z))]
\label{cor}
\end{eqnarray}
can be formulated.
In Eq.~(\ref{cor}), $f_1(z)$ describes the inhomogeneity of the
turbulent diffusivity, and we define $ f(z) = \alpha^{v}(z) f_1(z)$.
The profile $f(z)$ depends on details of the galactic structure.
Also, $\alpha (B)$ is the total $\alpha$ effect and $ \alpha = \alpha
^v \phi (B) + \alpha^h \phi_{1} (B)$, where $B=|{\bf B}|$.
Here $\alpha^v$ is the hydrodynamic part of the $\alpha$ effect,
with $ \alpha ^v \phi (B)$ its modification due to nonlinear
effects. Correspondingly, $\alpha^h$ is the magnetic part of the $\alpha$ effect,
and $ \alpha ^h \phi_{1} (B) $ is the modification caused by
nonlinear effects (see Rogachevskii and Kleeorin, 2000).
$\phi_{1} (B) = (3 / 8 B^{2}) (1 - \arctan (\sqrt{8} B) / \sqrt{8} B)
$ and the function $ \phi (B) $ is defined below.
The magnetic part of the $\alpha$ effect is proportional to the
magnetic helicity, i.e., $\alpha^h = \chi^{h} / (18 \pi \eta_{T}
\rho) $ (see, e.g., Kleeorin and Rogachevskii, 1999).
For galaxies the term $\alpha^h / T$ is very small and can
be dropped. The gauge conditions
$\bec{\nabla} \cdot {\bf A} = \bec{\nabla} \cdot {\bf a} = {\bf 0}$
have been used; our results can be shown to be gauge invariant
(see Berger and Ruzmaikin, 2000).

The last term in Eq.~(\ref{cor}) is related to the turbulent flux of
magnetic helicity. This turbulent flux is proportional to the
hydrodynamic part of the $\alpha$ effect and the turbulent diffusivity
(see Kleeorin and Rogachevskii, 1999). The turbulent flux of magnetic
helicity serves as an additional nonlinear source in the equation for
the magnetic part of the $\alpha$ effect and it causes a drastic
change in the dynamics of the large-scale magnetic field.

For simplicity we replace the flux divergence in the right hand side of
Eq.~(\ref{cor}) by a decay term, i.e. we replace ${\partial} \over
{\partial z}$ by $1/h$ (in principle, there is no problem in treating
this point more carefully).

\section{The equilibrium magnetic field configuration}

We now present some asymptotic expansions for galactic dynamo
models with the nonlinearity (\ref{cor}).
First of all, we recognize that, because of the large parameter
$4(h/l)^2$ in the right hand side of Eq.~(\ref{cor}), we can take

\begin{equation}
\alpha (B) = f(z) \phi (B) + R_\alpha ^{-1} B^{-2}{\bf B \cdot curl
\, B}, \label{alpha}
\end{equation}
where
\begin{eqnarray*}
\phi (B) &=& {3 \over 14 B^{2}} \biggl(1 - {\arctan (\sqrt{8} B)
\over \sqrt{8} B} + 2 B^{2} [ 1
\\
&-& 16 B^{2} + 128 B^{4} \ln (1 + (8 B^{2})^{-1})] \biggr).
\end{eqnarray*}
Thus
$\phi (B) = 1/(4B^2)$ for $B \gg 1/\sqrt{8}$ and $\phi=1-(48/5)B^2$
for $B \ll 1/\sqrt{8}.$ The function $\phi$ is derived by
Rogachevskii and Kleeorin (2000). Note that in a more simplified
model of turbulence the function $\phi (B) = \phi_{1} (B) = (3 / 8
B^{2}) (1 - \arctan (\sqrt{8} B) / \sqrt{8} B) $ (see Field et al.
1999).
We stress that the qualitative behaviour of the model does not depend
on these uncertainties in estimates for the  scaling functions $\phi$
and $\phi_1$.

Now we insert the $\alpha$-coefficient given by Eq.~(\ref{alpha})
into local
disc dynamo problem  to obtain the following equations:

\begin{eqnarray}
{{\partial b_r} \over {\partial t}} &=& - (\alpha (B) B_\phi)' +
b_r'',
\label{B6}\\
{{\partial B_\phi} \over {\partial t}} &=& D b_r + B''_\phi
\label{B7}
\end{eqnarray}
(here $B_r=R_\alpha b_r$). We can then obtain the steady-state
solution of
Eqs. ~(\ref{B6}) and (\ref{B7}).
Recognizing that in cylindrical coordinates
\begin{eqnarray}
{\bf B \cdot curl \, B} = R_\alpha (B_\phi b'_r - b_r B'_\phi),
\label{A3}
\end{eqnarray}
we obtain for fields of
quadrupole symmetry (cf. Kvasz et al., 1992)
\begin{eqnarray} B'''_\phi + D \alpha (B) B_\phi = 0
\label{newBB}
\end{eqnarray}
in a steady state. The corresponding equation in kinematic theory
reads
\begin{equation}
B'''_\phi + D \alpha_{0} B_\phi = 0.
\label{A4}
\end{equation}
Substituting (\ref{alpha}) into (\ref{newBB}) we obtain,
\begin{equation} B'''_\phi B^2 + D B_\phi [f(z) \phi(B) B^2 +
R_\alpha ^{-1} {\bf B \cdot curl \, B}] = 0 .
\label{newBD}
\end{equation}
Using Eq.~(\ref{A3}) we rewrite Eq.~(\ref{newBD}) in the form
\begin{equation} B'''_\phi (B^2 - B_\phi^{2}) + B_\phi [B''_\phi
B'_\phi + D f(z) \phi(B) B^{2}] = 0 .
\label{newDD}
\end{equation}
For the $\alpha \Omega$ dynamo $ B \approx B_\phi .$ This assumption
is justified if $|D|>>R_\alpha$, i.e. $|R_\omega|>>1$.
Eq.~(\ref{newDD}) then becomes
\begin{equation}
B'' B' + D f(z) \phi(B) B^{2} = 0,
\label{B8}
\end{equation}
Note that Eq.~(\ref{B8}) differs from Eq.~(\ref{A4}), arising from
kinematic theory.
For the specific choice of helicity profile $f(z)=\sin\pi z$ and
negative dynamo number $D$, there is an explicit steady solution, if
we assume $B^2\approx B_\phi^2$
(remember that also $B \gg 1/\sqrt{8}$, i.e.
super-equipartition), of the form
\begin{eqnarray}
B_\phi &=& {{2 \sqrt {|D|}} \over {\pi^{3/2}}} B_{\rm eq} \cos \,
{{\pi z} \over 2},
\label{final}\\
B_r &=& - {\sqrt{\pi R_\alpha} \over 2 \sqrt {|R_\omega|} } B_{\rm
eq} \cos \, {{\pi z} \over 2},
\label{finalr}
\end{eqnarray}
where we have restored the dimensional factor $B_{\rm eq}$.
(Note that ${\bf B \cdot curl \, B}=0$ for this approximate
solution.) This solution is remarkably close to the results from the
naive {\it Ansatz} $\alpha = \alpha_0 (1 - (B/B_{\rm eq})^2)$ or
$\alpha = \alpha_0 /(1+ (B/B_{\rm eq})^2)$, or the model of Moss et
al. (1999).
For example, the pitch angle of the magnetic field lines is $p= -
{\rm arctan} \, (\pi^2/4|R_\omega|) \approx 14^\circ$ for $|D| = 10$
and $R_\alpha = 1$.

\section{Numerical results}

We verified numerically that the initially weak magnetic field
approaches the equilibrium configuration (\ref{final}) with accuracy
1\% for $|D| > 1000$, and an accuracy of $50\%$ for $|D|>10$.
As is anticipated in the
previous section, the equilibrium magnetic field near to the
generation threshold value is more complicated. The threshold value
for the nonlinear solution of Eqs. (\ref{B6}) and (\ref{B7}) is $D
\approx - 3.14$, while the linear threshold value is $D \approx -8$.
This is because the nonlinear solution arranges itself so that the
term ${\bf B \cdot curl \, B}/ B_\phi ^2$ in $\alpha$ (see
Eq.~(\ref{alpha})) is of order 1. Thus, for the nonlinear solution
with $D = -8$, the maximal value of $\alpha$ is about 1.25, whereas
for $D = -5$, the maximal value is about 1.76. For $|D| \lta 10$ we
obtain numerically \begin{equation} B_\phi (0) \approx 0.23 |D-D_{\rm
cr}|^{0.52}, \label{lownon} \end{equation} where $D_{\rm cr}$ is the
nonlinear threshold value.
As $|D-D_{\rm cr}|$ increases towards 10, the slope increases
slightly, but Eq. (\ref{lownon}) remains a reasonable estimate.
(Note that accurately estimating the exponent in Eq.~(\ref{lownon}),
and subsequently, is a quite delicate matter even in this one-dimensional
problem, and that the quoted figures may be uncertain in the last digit.)

This result is robust under variations of the helicity profile. For $f(z)
=z$ we get in the nonlinear case $D_{\rm cr} = -7.49$, while the
linear threshold value is $D_{\rm cr} = -12.5$ and $B_\phi (0)
\approx 0.15 |D-D_{\rm cr} |^{0.50}$ near $D=D_{\rm cr}$, i.e. again a
square root dependence to within the errors of our procedure.
Further, with $f(z)= z/|z|$ in $|z|>0.2$, and a smooth interpolation
to zero in $|z|\leq 0.2$, we find $D_{\rm cr}\approx -2.41$ and
$B_\phi (0) \approx 0.25 |D-D_{\rm cr}|^{0.50}$, again closely the
same dependence. (In this case the linear threshold value is $D_{\rm
cr}=-6.53$.)

\section{Discussion}

We have demonstrated that the nonlinear evolution of the helicity
following from Eq.~(\ref{prime}) gives a basically different type of
galactic magnetic field evolution to that following from
Eq.~(\ref{cor}). Eq.~(\ref{prime}),
being based on local helicity conservation, results in magnetic field
decay, after a stage of kinematic growth. If the molecular
diffusivity of the magnetic field is taken into account, this decay
is followed by a stabilization at a very low magnetic field strength,
corresponding to the estimate of Vainshtein and Cattaneo (1992). The
scenario of magnetic field and helicity dynamics can then be
described as follows. Large-scale dynamo action produces large-scale
magnetic helicity. Due to the local conservation of helicity,
suppression of field generation results. An equilibrium is possible
if molecular diffusivity is present, so the equilibrium magnetic
field strength is very low.

Equation ~(\ref{cor}) allows for the transport of helicity, so the
local value of the helicity changes during magnetic field evolution.
The scenario of magnetic field and helicity dynamics can be presented
as follows. As usual, magnetic helicity of the large-scale magnetic
field is produced, however the total magnetic helicity is not now
conserved locally, but the magnetic helicity of the small-scale
magnetic field is redistributed by a helicity flux.  The equilibrium
state is given by a balance between helicity production and
transport. The helicity conservation law now expresses the
conservation of an integral of the helicity over the galactic disc.
However this conservation law is trivial, because the integral
vanishes identically as helicity is an odd function with respect to
$z$.
Now the equilibrium strength of the large-scale magnetic field is of
order that of the equipartition field: this is our main result.

\begin{acknowledgements}
We acknowledge support from INTAS Program Foundation (Grant No.
99-348),
NATO Collaborative Linkage Grant PST.CLG~974737 and RFBR grant
00-05-64062a.
DS is grateful to the Royal Society for financial support and to a
special
fund of the Faculty of Engineering of the Ben-Gurion University of
the Negev
for visiting senior scientists. We thank A.Brandenburg
for helpful comments.
\end{acknowledgements}


\begin{thebibliography}{}

\bibitem[1996]{}
Beck R., Brandenburg A., Moss D., Shukurov A., Sokoloff D., 1996,
Ann. Rev. Astron. Astrophys. {\bf 34}, 155

\bibitem[2000]{}
Berger M.A., Ruzmaikin A., 2000, J. Geoph. Res., in press

\bibitem[1999]{}
Blackman E.G., Field G.B., 1999, ApJ {\bf 521}, 597

\bibitem[2000]{}
Blackman E.G., Field G.B., 2000, ApJ {\bf 534}, 984

\bibitem[2000]{}
Brandenburg A., 2000, astro-ph/0006186

\bibitem[1997]{}
Brandenburg A., Donner K.-J., 1997, MNRAS {\bf 288}, L29

\bibitem[]{}
Brandenburg A., Zweibel E., 1995, ApJ {\bf 448}, 734

\bibitem[]{}
Field G., Blackman E., Chou H., 1999, ApJ {\bf 513}, 638

\bibitem[1996]{}
Cattaneo F., Hughes D., 1996, Phys. Rev. E {\bf 54} R4532

\bibitem[1995]{}
Gruzinov A.V., Diamond P.H., 1995, Phys. Plasmas {\bf 2}, 1941

\bibitem[1982]{}
Kleeorin N.I., Ruzmaikin A.A., 1982, Magnetohydrodyn. {\bf 18}, 116

\bibitem[1999]{}
Kleeorin N., Rogachevskii I., 1999, Phys. Rev. E. {\bf 59}, 6724

\bibitem[1995]{}
Kleeorin N., Rogachevskii I., Ruzmaikin A., 1995, Astron. Astrophys.
{\bf 297}, 159

\bibitem[1999]{}
Kulsrud R., 1999, Ann. Rev. Astron. Astrophys. {\bf 37}, 37

\bibitem[1992]{}
Kvasz L., Shukurov A., Sokoloff D., 1992, Geophys. Astrophys. Fluid
Dyn.
{\bf 65}, 231

\bibitem[1999]{}
Moss D., Shukurov A., Sokoloff D., 1999, Astron. Astrophys. {\bf
343}, 120

\bibitem[1992]{}
Parker E., 1992, ApJ {\bf 401}, 137

\bibitem[2000]{}
Rogachevskii I., Kleeorin N., 2000,  Phys. Rev. E. {\bf 61}, 5202

\bibitem[1988]{}
Ruzmaikin A., Shukurov A., Sokoloff D., 1988, Magnetic Fields of
Galaxies.
Kluwer, Dordrecht

\bibitem[1992]{}
Vainshtein S.I., Cattaneo F., 1992, ApJ {\bf 393}, 165

\bibitem[1983]{}
Zeldovich Ya.B., Ruzmaikin A.A., Sokoloff D.D., 1983, Magnetic Fields
in Astrophysics. Gordon \& Breach, New York

\end{thebibliography}
\end{document}